\documentclass[aps, prd, twocolumn, superscriptaddress, longbibliography]{revtex4-2}
\usepackage{amsmath}
\usepackage{amssymb} 
\usepackage{graphicx}
\usepackage{enumerate}
\usepackage{color}
\usepackage{tensor}
\usepackage{mathrsfs}
\usepackage{hyperref}
\usepackage[normalem]{ulem}
\usepackage[dvipsnames]{xcolor}
\usepackage{accents}
\usepackage{scalerel}
\usepackage{bm}
\usepackage{anysize}
\usepackage{siunitx}

\usepackage{multirow}

\makeatletter
\newcommand*{\leqdef}{\mathrel{\rlap{%
			\raisebox{0.25ex}{$\m@th\cdot$}}%
		\raisebox{-0.25ex}{$\m@th\cdot$}}%
	=}
\newcommand*{\reqdef}{=\mathrel{\rlap{%
			\raisebox{0.25ex}{$\m@th\cdot$}}%
		\raisebox{-0.25ex}{$\m@th\cdot$}}
}
\makeatother

\begin{document}
\title{New dynamical system approach to Palatini \texorpdfstring{$f(R)$}{fR} theories and its application to exponential gravity}

\author{Jo\~ao C. Lobato}
\email{jcavlobato@if.ufrj.br}

\author{Isabela S. Matos}
\email{isa@if.ufrj.br}

\author{Maur\'\i cio O. Calv\~ao}
\email{orca@if.ufrj.br}

\author{Ioav Waga}
\email{ioav@if.ufrj.br}

\affiliation{Universidade Federal do Rio de Janeiro\\
	Instituto de F\'\i sica\\
	CEP 21941-972 Rio de Janeiro, RJ, Brazil}

\begin{abstract}
	The approach of dynamical systems is a useful tool to investigate the cosmological history that follows from modified theories of gravity. It provides qualitative information on the typical background solutions in a parametrized family of models, through the computation of the fixed points and their characters (attractor, repeller or saddle), allowing, for instance, the knowledge of which regions on the parameter space of the models generate the desired radiation, matter and dark energy dominated eras. However, the traditional proposal for building dynamical systems for an $f(R)$ theory in the Palatini formalism assumes the invertibility of a function that depends on the specific Lagrangian functional form, which is not true, for example, for the particular theory of exponential gravity ($f(R)=R-\alpha R_*(1-e^{-R/R_*})$). In this work, we propose an alternative choice of variables to treat  $f(R)$ models in their Palatini formulation, which do include exponential gravity. We derive some general results that can be applied to a given model of interest and present a complete description of the phase space for exponential gravity. We show that Palatini exponential gravity theories have a final attractor critical point with an effective equation of state parameter $w_{\text{eff}} = -1$ (for $\alpha>1$), $w_{\text{eff}} = -2/3$ (for $\alpha=1$) and $w_{\text{eff}} = 0$ (for $\alpha<1$). Finally, our analytical results are compared with numerical solutions of the field equations.
	\end{abstract}

\maketitle
\section{Introduction}
\label{sec:intro}
One of the greatest challenges in modern cosmology is to identify the physical mechanism responsible for the late-time cosmic acceleration. The two main theoretical
approaches to explain cosmic acceleration are the following: (1) assume the existence of an
unknown component with sufficiently negative pressure, generically denominated ``dark energy'', and (2) ``modified gravity'' in which general
relativity (GR) is modified at large scales or, more accurately, at low curvatures. The simplest dark energy candidate is Einstein's cosmological
constant ($\Lambda $). Although in very good accordance with current cosmological observations, $\Lambda$ faces some theoretical difficulties, such as its tiny value when comparing the theoretical expectation to the vacuum energy density, the so-called cosmic
coincidence and its fine-tuning. This situation has motivated the search for
alternatives like modified gravity theories. The simplest modified-gravity
candidate is the so-called $f(R)$ gravity in which the Lagrangian (density) is
a nonlinear function of the Ricci scalar $R$.

To obtain the field equations in $f(R)$ theories, two main variational
approaches can be adopted, namely, the metric or the Palatini formalisms.
The distinction lies on which gravitational (geometric) fields are considered as independent:
only the metric in the former and both the metric and the connection in the latter.
For the standard Einstein-Hilbert gravitational Lagrangian (with or without a cosmological constant), assuming that the  matter Lagrangian does not depend on the connection, both approaches lead to the same field equations. However, for a  nonlinear $f(R)$, the two methods give rise to different field equations and distinct cosmical dynamics; in fact, in the general case, the metric approach
yields fourth-order field equations, while the Palatini approach generates a second-order system and is, therefore, more easily tractable. An important feature of a large class of $f(R)$ gravity theories is that an accelerated expansion appears naturally in both methods. 

In this work, we perform a careful dynamical system analysis of $f(R)$ under the Palatini formalism \cite{Olmo2011} and apply it to the special case of exponential gravity theory \cite{Linder2009,Cognola2008}.
Dynamical system analysis has been explored with great success in the study of several cosmologies \cite{Bogoyavlensky1985,Wainwright1997,Coley2003,Bahamonde2018}.
Of special interest here is the work of reference \cite{Fay2007} (hereafter called FTT) that analyzed the cosmological viability of $f(R)$ theories under the Palatini approach. 
They investigated the possibility of cosmologies having four relevant phases: early inflation, radiation and (non-relativistic) matter dominated eras and a late-time accelerated expansion.  
To this end they considered, for instance, the case of power laws of the type $f(R)=R+\alpha R^m-\beta/R^n$ ($\alpha, \beta>0$). If the term $\alpha R^m$ is dominant at relatively large curvatures, in principle, it could drive early-inflation, while if the term $\beta/R^n$ is dominant at low curvatures, depending on the value of $n$, it could drive late-time acceleration. 
FTT showed that for this $f(R)$ theory an early inflationary era is not compatible with a subsequent standard radiation-dominated era. 
They remarked that although a sequence of four phases are not possible for the above model, three out of four are possible.
Here we are mainly interested in Palatini exponential gravity and, in our case, besides matter and radiation dominated eras, only late-time cosmic acceleration is expected to be relevant.

As will become clear later on, Palatini exponential gravity cannot fully be treated by using the FTT formalism because of inversion problems. As will be shown, in order to have a closed autonomous system on the FTT approach, there is a function of the Ricci scalar, $C(R)$, in the dynamical system of equations, that needs to be expressed in terms of the FTT original variables $\tilde{y}_1$ and $\tilde{y}_2$. This cannot be done fully in the Palatini exponential gravity theory.  In order to completely analyze this theory, it is necessary to introduce new variables $y_1$ and $y_2$. Therefore, in this sense, the FTT Palatini approach cannot generically be applied to all $f(R)$. For the Palatini exponential gravity the formalism we propose corrects this limitation. Similar inversion problems as the one that will be responsible for the failure of the FTT approach on the Palatini exponential gravity are discussed and solved in the metric formalism for a given $f(R)$ in \cite{Carloni2015}.

This paper is organized as follows. Section \ref{sec:FTT} presents the traditional FTT approach, its main results and limitations, motivating the need of new variables for the study of the exponential gravity theory. In Section \ref{sec:new_variables} the new variables are introduced together with the dynamical system equations written in terms of them. In Section \ref{sec:results} the main results are obtained for a general $f(R)$ regarding the critical points of the system and their nature and a more detailed analysis is made for the exponential gravity theory at the end. Finally, Section \ref{sec:discussion} discusses the results obtained, comparing it with what one would obtain in the FTT approach in the parameter regime where it is valid. Numerical particular solutions are presented as well to exemplify qualitative behaviors foreseen by the dynamical system analysis.

Our sign conventions are those of \cite{Misner1973}, and we use units such that the vacuum speed of light is $c=1$ and thus the Einstein gravitational constant is $\kappa\leqdef8\pi G_N$\,.

\section{Traditional approach to Palatini \texorpdfstring{$f(R)$}{fR} dynamical systems} \label{sec:FTT}

\subsection{Palatini \texorpdfstring{$f(R)$}{fR} theories}

In the usual first-order Palatini variational approach to $f(R)$ modified theories of gravity \cite{Hamity1993,Ferraris1994, Vollick2003,Olmo2011}, three sets of independent fields are considered: (i) the matter fields $\psi_A$ (where $A$ is a collective index taking into account all kinds of non-gravitational fields), (ii) the metric tensor $g^{\alpha\beta}$, and (iii) the (affine) connection $\Gamma^\alpha_{\mu\nu}$, where the last two stand for the gravitational fields (in contrast to the Einstein-Hilbert approach). The total action is given by
\begin{align}
    S &\leqdef S_G + S_M\,, \\
    \intertext{where}
    S_G[g^{\alpha\beta}, \Gamma^\alpha_{\mu\nu}, \Gamma^\alpha_{\mu\nu,\beta}] &\leqdef-\frac{1}{2\kappa}\int_{\mathcal{M}}f(R)\sqrt{-g}\,d^4x\,,
       \intertext{and}
    S_M[g^{\alpha\beta}, \psi_A, \psi_{A,\alpha}]&\leqdef\int_{\mathcal{M}}L_M\sqrt{-g}\,d^4x\,.
\intertext{Here the Ricci scalar is defined by}
    R&\leqdef g^{\mu\nu}R_{\mu\nu}\,,
\intertext{whereas the (symmetric) Ricci tensor is the usual function of the connection only:}
    R_{\mu \nu } \leqdef \Gamma _{\mu \nu,\alpha }^{\alpha
} & - \Gamma _{\mu \alpha,\nu }^{\alpha }+\Gamma _{\sigma \alpha
}^{\alpha }\Gamma _{\mu \nu }^{\sigma }-\Gamma _{\sigma \nu }^{\alpha
}\Gamma _{\mu \alpha }^{\sigma }\,,
 \end{align}
and $L_M(g^{\alpha\beta}, \psi, \psi_{A,\alpha})$ is the matter Lagrangian.

When extremizing $S$ with respect to $g^{\mu\nu}$, we get
\begin{equation}
\label{field_eq}
    f'R_{\mu\nu} - \dfrac{1}{2}f g_{\mu\nu} = \kappa T_{\mu\nu}\,,
\end{equation}
where, as usual (since the $L_M$ does not depend on $\Gamma^\alpha_{\mu\nu}),$ the energy-momentum tensor (EMT) is given by
\begin{equation}
    T_{\mu\nu}\leqdef \dfrac{2}{\sqrt{-g}}\dfrac{\delta(\sqrt{-g}L_M)}{\delta g^{\mu\nu}}\,.
\end{equation}
Here, of course, $f'\leqdef df/dR$\,.

When extremizing $S$ with respect to $\Gamma^\alpha_{\mu\nu}$, the corresponding equations are equivalent to
\begin{equation}
    \nabla_\alpha\left(f'\sqrt{-g}g^{\mu\nu} \right)=0\,,
\end{equation}
where $\nabla_\alpha$ is the covariant derivative operator associated to $\Gamma^\alpha_{\mu\nu}$\,. This equation may be used in order to express the connection in terms of $f'$ and $g^{\mu\nu}$ and it turns out that $\Gamma^\alpha_{\mu\nu}$ are the Christoffel symbols associated to the non-degenerate symmetric tensor $h_{\mu\nu}\leqdef f'g_{\mu\nu}$ or, equivalently, 
\begin{align}
    \Gamma _{\mu \nu }^{\alpha } = \left\{ _{\mu \nu }^{\alpha }\right\} +\frac{1}{%
2f'}\left[2\delta _{(\mu }^{\alpha }\partial _{\nu )}f'-g^{\alpha
\sigma }g_{\mu \nu }\partial _{\sigma }f'\right]\,, \label{connection}
\end{align}
where $\left\{ _{\mu \nu }^{\alpha }\right\}$ are the Christoffel symbols of the metric $g_{\mu \nu}$.

Finally, for our specific total action, one can also show \cite{Koivisto2006} that the EMT obeys the usual conservation law:
\begin{equation}
    {T^{\mu\nu}}_{;\nu} = 0\,, \label{EMTconservation}
\end{equation}
where the subindex ; denotes the covariant derivative with respect to the metric connection.

Considering from now on a spatially flat Friedmann-Lemaître-Robertson-Walker (FLRW) metric,
\begin{align}
ds^2 = -dt^2 + a(t)^2\left(dx^2 + dy^2 + dz^2\right)\,, \label{metric}
\end{align}
we assume the EMT to be composed of regular radiation and non-relativistic matter, that is
\begin{align}
T_{\mu \nu} = \left(\rho_m + \frac{4}{3}\rho_r\right)u_ {\mu} u_{\nu} + \frac{1}{3}\rho_r g_{\mu \nu}\,, \label{EMT}
\end{align}
where $\rho_m$ and $\rho_r$ are the matter and radiation energy densities, respectively, and $u_{\mu}$ is the Hubble flow four-velocity. We will assume these components do not interact and thus Eq. (\ref{EMTconservation}) leads, separately, to the familiar expressions
\begin{align}
\rho_m = \rho_{m0}\text{e}^{-3N}\,, \label{matter_conservation}\\
\rho_r = \rho_{r0}\text{e}^{-4N}\,,\label{radiation_conservation}
\end{align}
where we assume $\rho_{m0}$ and $\rho_{r0}$ as positive quantities throughout this work and anticipate the convenience of using as independent variable the e-fold parameter (instead of the scale factor or redshift)
\begin{equation}
N \leqdef \ln a\,.
\end{equation}

One difference between the metric and Palatini formalisms regard their equivalent representation as a scalar-tensor theory. Both can be treated as a Brans-Dicke theory with a potential \cite{Olmo2011}, but the former has a Brans-Dicke parameter equal to 0, while the latter has the same parameter equal to $-3/2$. The main physical consequence of that difference is that on the metric formalism, the scalar field introduces an additional dynamical degree of freedom when compared with general relativity, while in the Palatini case, it can be shown that the scalar field satisfies an algebraic relation, not having the same dynamical behavior. 

Another fundamental difference between both approaches that is worth stressing again is regarding the order of the field equations. In the metric formalism a set of fourth-order differential equations on the metric are found, whereas in the Palatini case the equations are of second order. This difference simplifies the trace of the generalized Einstein's equations, Eq.~(\ref{field_eq}), which becomes a purely algebraic relation between $\rho_m$ and $R$:
\begin{align}
\kappa \rho_{m} = 2f - f'R\,.  \label{trace}
\end{align}

Writing the Ricci scalar with the help of Eq.~(\ref{connection}) and substituting it in  Eq.~(\ref{field_eq}) together with Eq.~(\ref{EMT}), one obtains \cite{Olmo2011} the generalized Friedmann equation,
\begin{align}
6f'H^{2}\left( 1+\frac{f''}{2f'}\frac{dR}{dN}\right)
^{2}-f =\kappa (\rho_{m}+2\rho_{r}),
\label{Friedmann}
\end{align}
where, of course, $H\leqdef \dot{a}/a$. By differentiating Eq.~(\ref{trace}), then using Eqs.~(\ref{matter_conservation}) and (\ref{trace}), we find
\begin{align}
\frac{dR}{dN} = -3\frac{f'R - 2f}{f''R - f'}\,, \label{ricci_evolution}
\end{align}
from which we can rewrite Eq.~(\ref{Friedmann}) as
\begin{align}
H^{2} = \frac{2\kappa (\rho_{m} + \rho_{r})+ f'R-f}{6f'\xi }
,  \label{Hubble_Param}
\end{align}
with
\begin{align}
\xi \leqdef \left[ 1-\frac{3f''(f'R-2f)}{%
	2f'(f''R-f')}\right] ^{2}.
\end{align}

Finally, differentiating Eq.~(\ref{Hubble_Param}) and using Eqs.~(\ref{matter_conservation}), (\ref{radiation_conservation}) and (\ref{Hubble_Param}), we obtain
\begin{align}
	 \frac{dH^2}{dN}  = -3H^2 + 3\frac{f'R - f}{6f'\xi} - \frac{\kappa \rho_r}{3f'\xi}&\nonumber \\ - \frac{\dot{f}'H}{f'} - \frac{\dot{\xi}H}{\xi}+\frac{\dot{f}'R}{6f'\xi H}&\,, \label{Hubble_derivative}
\end{align}
where $\, \dot{} \leqdef d/dt = Hd/dN$.
 
We point out that one can write $\rho_r$ only in terms of $H$ and $R$ by replacing Eq.~(\ref{trace}) in Eq.~(\ref{Hubble_Param}) and then rewrite the evolution of the Hubble parameter as
\begin{align}
 \frac{dH^2}{dN}  = & -4H^2 + \frac{R}{3\xi} \nonumber \\ &+ \frac{3H^2(f'R - 2f)}{2(f''R - f')}\left[ \frac{f''}{f'} + \frac{\xi'}{\xi}-\frac{f''R}{6f'\xi H^2}\right]. \label{Hubble_evolution}
\end{align}
Then, Eqs.~(\ref{ricci_evolution}) and (\ref{Hubble_evolution}) constitute an autonomous system of differential equations for the variables $H^2$ and $R$, with the independent parameter given by $N$. All the formalism of dynamical systems can then be employed when analyzing the qualitative behavior of the solutions near the critical points of the theory. However, this is not suitable for two reasons: in these variables,  we expect the existence of critical points in the infinity ($R \rightarrow \infty$ or $H \rightarrow \infty$) and, as we will see later, a pair of values of $R$ and $H$ does not completely specify a physical cosmological solution. Therefore, it is often convenient to introduce new variables in order to simplify the analysis and avoid these subtleties, specially when the intention is to describe several $f(R)$ models at once, as we shall see next.

\subsection{FTT approach}

The traditional way of addressing the dynamical system in the Palatini formalism for $f(R)$ gravity theories is found in \cite{Fay2007}. Here we summarize the procedure. First, we express the dynamical system in terms of the variables:
\begin{align}
	\bar{y}_1 \leqdef  \frac{f'R - f}{6f'\xi H^2}\,, \quad
	\bar{y}_2 \leqdef  \frac{\kappa \rho_r}{3f'\xi H^2}\,. \label{FTT}
\end{align}
Applying the derivative to both of these expressions and using Eqs.~(\ref{radiation_conservation}), (\ref{ricci_evolution}) and (\ref{Hubble_evolution}), we arrive at an equivalent dynamical system given by:
\begin{align}
	&\frac{d\bar{y}_1}{dN} = \bar{y}_1 [3 - 3\bar{y}_1 + \bar{y}_2 + C(R)(1-\bar{y}_1)]\,, \label{y1FTT_evolution} \\
	& \frac{d\bar{y}_2}{dN} = \bar{y}_2[-1-3\bar{y}_1+\bar{y}_2-C(R)\bar{y}_1]\,,\label{y2FTT_evolution}
\end{align}
where 
\begin{align}
	C(R)\leqdef -3 \frac{(f'R - 2f)f''R}{(f'R-f)(f''R - f')}\,. \label{Cdef}
\end{align}
The solutions of the system are the curves $(\bar{y}_1(N),\bar{y}_2(N))$ on the phase space of the points $(\bar{y}_1,\bar{y}_2)$. The critical points of the dynamical system are obtained by setting $d\bar{y}_1/dN = d\bar{y}_2/dN = 0$ and solving for $\bar{y}_1$ and $\bar{y}_2$. Any solution that starts at a critical point remains on it. Depending on the critical point nature, solutions nearby can be attracted or repelled by it. To study which behavior occurs, one linearizes the right-hand-side of the two differential equations of the system and find the eigenvalues of the corresponding Jacobian matrix evaluated at each invariant point.  If both eigenvalues are negative, the critical point is an attractor and all neighboring solutions tend to evolve into it. If both are positive, all solutions on its vicinity are repelled. Finally, if one eigenvalue is positive and the other is negative, one has a saddle point that attracts solutions in some directions and repels in others.

Assuming $C(R) \neq -3, -4$, one concludes that the critical points of the Palatini $f(R)$ theories of gravity are given in the $(\bar{y}_1,\bar{y}_2)$ plane by:
\begin{itemize}
	\item $P_r \leqdef (0,1)$ with eigenvalues $\{4+C(R),1\}$, 
	\item $P_m \leqdef (0,0)$ with eigenvalues $\{3+C(R),-1\}$, 
	\item $P_d \leqdef (1,0)$ with eigenvalues $\{-3-C(R),$ $-4 - C(R)\}$.
\end{itemize} 

Although this procedure can be applied for a variety of $f(R)$ theories in the Palatini formalism, it is not well defined in some cases, since Eqs. (\ref{y1FTT_evolution}) and (\ref{y2FTT_evolution}) are not always entirely expressible in terms of $\bar{y}_1$ and $\bar{y}_2$. For this to be possible, and the system recognized as a true closed dynamical one, it is necessary to express $C(R)$ as a function of these variables. This can be achieved by additionally imposing Eq.~(\ref{Hubble_Param}), which is now a constraint equation that can be written, considering Eq.~(\ref{trace}), as
\begin{align}
	\frac{f'R-2f}{f'R - f} = \frac{\bar{y}_1 + \bar{y}_2 - 1}{2\bar{y}_1}. \label{FTT_constraint}
\end{align}

For a given $f(R)$, the above equation can, in principle, be inverted to provide $R$ as a function of the variables $\bar{y}_1$ and $\bar{y}_2$. The function $R(\bar{y}_1, \bar{y}_2)$  can then be replaced in $C(R)$, assuring the closed character of the system. An exception to this procedure is when the LHS of the above equation is a non-invertible function of $R$. When this occurs, the same values of $(\bar{y}_1, \bar{y}_2)$ may correspond to multiple values of $R$ and $H$, leading to the overlap of different critical points in the phase space of these variables and precluding a clear description of the behavior of the solutions near them. Such inversion problem happens, for instance, in exponential gravity, defined by
\begin{align}
f(R) = R - \alpha R_{\ast}(1 - e^{-R/R_{\ast}})\,, \label{exp_gravity}
\end{align}
where $R_{\ast}$ and $\alpha$ are free positive parameters. In this case, when $\alpha<1$, the system cannot be closed and a new set of variables is needed.

The explicit inversion of Eq.~(\ref{FTT_constraint}) is particularly important when $C(R)$ assumes the values $-3$ or $-4$. This is because there can be extra critical points, as long as they satisfy:
\begin{itemize}
    \item $(\bar{y}_1, \bar{y}_2) = (0, 1)$ or $(\text{const}, 0)$ and $C\left(R(\bar{y}_1, \bar{y}_2)\right)= -3\,$,
    \item $(\bar{y}_1, \bar{y}_2) = (0, 0)$ or $(1, 0)$ or ($\text{const}, 1 - \text{const})$ and $C\left(R(\bar{y}_1, \bar{y}_2)\right)= -4\,$.
\end{itemize}

Therefore, despite obtaining quite general results, it is not clear from this approach whether a generic $f(R)$ indeed contains the three desired radiation/matter/dark energy dominated phases. Besides, when $\bar{y}_1 = 0$, the inversion of the constraint is not possible as well.

\section{New variables}
\label{sec:new_variables}
In order to deal with some theories in which the traditional approach is problematic, we propose a new set of variables:
\begin{align}
y_1 \leqdef \frac{f}{2H^2f'\xi}\,, \quad y_2 \leqdef y_1 + \frac{f}{R}.\label{new_variables_def}
\end{align}
Whenever $f/R$ is an invertible function of $R$, the second definition can be used to obtain the Ricci scalar as a function of the variables $y_1$ and $y_2$. It is instructive to notice that, in the $(y_1, y_2)$ plane, the collection of points with the same value of $R$ is a straight line inclined by $45$ degrees and intercepting the $y_2$ axis in the point $(0, f/R)$, as indicated on Fig.~\ref{fig:phase_space}. 

In terms of the new variables, the constraining Eq.~(\ref{Hubble_Param}) can be expressed as
\begin{align}
\frac{\kappa \rho_r}{3H^2f'\xi} &= 1 - y_1 + \frac{y_1f'}{3(y_2 - y_1)}. \label{rho_r}
\end{align}
Differentiating Eq.~(\ref{new_variables_def}) and using Eqs.~(\ref{ricci_evolution}), (\ref{Hubble_evolution}) and (\ref{rho_r}), one arrives at the system
\begin{align}
\frac{dy_1}{dN} =& \frac{f' - 2(y_2 - y_1)}{f' - f''R}\frac{y_1}{y_2 - y_1}(3f' - y_1f''R) \label{y_1_evolution} \nonumber \\ &- \frac{2y_1^2f'}{3(y_2 - y_1)} + 4y_1\,, \\
\frac{dy_2}{dN} =& \frac{f' - 2(y_2 - y_1)}{f' - f''R}\bigg[\frac{3f'y_2 - y_1^2f''R}{y_2 - y_1} \nonumber \\ &- 3(y_2 - y_1)\bigg]  - \frac{2y_1^2f'}{3(y_2 - y_1)} + 4y_1 \label{y_2_evolution}
\end{align}
where all functions of $R$ are to be understood as functions of $y_1$ and $y_2$, since we assume that $R(y_1,y_2)$ can be obtained by the second definition in Eq.~(\ref{new_variables_def}). Under this condition, then, the system is closed.  

It is important to notice that our procedure is complementary to the traditional FTT one, since it also relies upon an invertibility assumption. Depending on which theory one wishes to study, one needs to choose the approach that suits better. It is, of course, also possible that neither of the approaches are viable, if both the LHS of Eq.~(\ref{FTT_constraint}) and $f/R$ are not invertible for $R$. Then, one will need to construct another set of variables to describe the system of interest, which can be made by simply exchanging $f/R$ for a chosen invertible function of $R$ in our definition of $y_2$. 

Furthermore, we can write the fractional energy densities of matter, radiation and of a geometric component, respectively, in terms of the new variables as:
\begin{align}
    &\Omega_m \leqdef \frac{\kappa \rho_m}{3H^2}  = \left[\frac{4y_1}{3} - \frac{2y_1f'}{3(y_2-y_1)} \right]f'\xi, \label{Omega_m}\\
    &\Omega_r\leqdef \frac{\kappa \rho_r}{3H^2} = \left[1- y_1+\frac{f'}{3(y_2-y_1)}y_1\right]f'\xi, \label{Omega_r}\\
    &\Omega_{\text{geo}} \leqdef 1 - \Omega_m - \Omega_r \label{Omega_de}.
\end{align}
The component related with the parameter $\Omega_{\text{geo}}$, as we will see, can be responsible for providing an accelerated expansion phase of the universe, being then called a dark energy parameter but it may as well behave effectively as dust or radiation in some cases. This will depend on the equation of state that the combined cosmic fluid will satisfy.
The effective equation of state parameter accounting for all components, defined by
\begin{align}
    \frac{\dot{H}}{H^2} \reqdef -\frac{3}{2}(1 + w_{\text{eff}})\,, \label{acceleration_eq}
\end{align}
can be written in terms of the new variables as
\begin{align}
w_{\text{eff}} =& -\frac{2y_1f'}{9(y_2 - y_1)} + \frac{1}{3} + \nonumber \\ &\frac{f' - 2(y_2 - y_1)}{f' - f''R}\left[\frac{f''R}{f'} + \frac{\xi'R}{\xi} - \frac{y_1f''R}{3(y_2 - y_1)}\right]. \label{weff}
\end{align}
It gives the equation of state of the total fluid,  establishing when it behaves effectively as dust, radiation or dark energy and determining the period of acceleration of the universe in a given solution of the system. In terms of the equation of state parameter of each component, the effective parameter is:
\begin{align}
    w_{\text{eff}} = \frac{\Omega_r}{3} + w_{\text{geo}}\Omega_{\text{geo}}. \label{individual_w}
\end{align}

\section{Critical points and phase space reconstruction} \label{sec:results}

We start by analyzing the case of a generic $f(R)$. We will only be interested in critical points where $y_1$ and $y_2$ are non-divergent. At a critical point, a necessary condition that comes from the derivative of the second definition in Eq.~(\ref{new_variables_def}) is:
\begin{align}
    \frac{d}{dN}\left(\frac{f}{R}\right) = -\frac{3(f'R - 2f)(f'R-f)}{(f''R - f')R^2} = 0\,. \label{critical_condition}
\end{align}
This implies that every critical point in which $f''R-f'\neq \pm \infty$ satisfies either
\begin{align}
    f' = \frac{2f}{R} = 2(y_2-y_1) \label{fixed_points_condition_1}
\end{align}
or
\begin{align}
    f' = \frac{f}{R} = y_2 - y_1. \label{fixed_points_condition_2}
\end{align}
By Eq.(\ref{trace}), the first of these conditions is equivalent to $\rho_m =0$. It is important to emphasize that some solutions of the above equations may still not correspond to a critical point and one has to verify if that is the case afterwards (\textit{i.e.} if $dy_1/dN = 0$ and $dy_2/dN = 0$). 

If $R$ is a solution of Eq.~(\ref{fixed_points_condition_1}), we find the corresponding value of $y_1$ by replacing this relation in the RHS of Eq.~(\ref{y_2_evolution}) and imposing a vanishing LHS. Assuming $f'-f''R \neq 0$ and $f''R^2/f\neq \pm \infty$ in Eq.~(\ref{y_2_evolution}), this leads to
\begin{align}
    y_1 = 3 \quad \text{or} \quad y_1 = 0\,. \label{y_1_first_case}
\end{align}
On the other hand, if $R$ satisfies  Eq.~(\ref{fixed_points_condition_2}), we replace this relation in the RHS Eq.~(\ref{y_1_evolution}) together with $dy_1/dN = 0$ and, assuming $f'-f''R \neq 0$ and $f''R \neq \pm \infty$ on it, we obtain
\begin{align}
    y_1 = \frac{3f' - 12f''R}{2f'-5f''R} \quad \text{or} \quad y_1 = 0\,. \label{y_1_second_case}
\end{align}
The corresponding values of $y_2$ are found by replacing the values of $y_1$ and $R$ into the second part of Eq.~(\ref{new_variables_def}). Note that each of Eqs.~(\ref{fixed_points_condition_1}) and (\ref{fixed_points_condition_2}) can have multiple solutions. Even though the Ricci scalar is assumed to be a function of $y_1$ and $y_2$, it might be the case that multiple solutions of $R$ are associated with a single value of $y_1$, but then they need to have distinct values of $y_2$. The $R$-dependent $y_1$ in Eq.~(\ref{y_1_second_case}) may, as well, have a different value for each of the solutions of Eq.~(\ref{fixed_points_condition_2}). If some value of $R$ that solves one of the Eqs.~(\ref{fixed_points_condition_1}) and (\ref{fixed_points_condition_2}) does not satisfy the appropriate above mentioned conditions (\emph{e.g.} if $f'-f''R=0$), a more careful limiting process must be done in Eqs.~(\ref{y_1_evolution}) and (\ref{y_2_evolution}) so that the critical points can be found. The result of this limit will depend, of course, on the functional form of $f(R)$.

We can already assess some aspects of the matter content and the acceleration character of the possible critical points. Assuming no divergence occurs and replacing Eqs.~(\ref{fixed_points_condition_1}) and (\ref{y_1_first_case}) in Eqs.~(\ref{Omega_m})--(\ref{Omega_de}) and (\ref{weff}):
\begin{align}
&\text{For } y_1 = 3: \;
    \begin{cases}
    \Omega_m = \Omega_r = 0, \; \Omega_{\text{geo}} = 1\,, \\
    w_{\text{eff}} = -1 \,. \label{Omegas_y1_3_first_case}
    \end{cases} \\
&\text{For } y_1 = 0    : \;
    \begin{cases}
    \Omega_m = 0,\text{ }  \Omega_{r} = 2y_2, \text{ } \Omega_{\text{geo}} = 1 - 2y_2\,, \\
    w_{\text{eff}} = \frac{1}{3}\,. \label{Omegas_y1_0_first_case}
    \end{cases}
\end{align}
When using Eq.~(\ref{fixed_points_condition_2}) in Eqs.~(\ref{Omega_m})--(\ref{Omega_de}) and (\ref{weff}), one finds:
\begin{align}
    \Omega_m &= \frac{2y_1(y_2-y_1)\xi}{3}\,, \nonumber\\
    \Omega_r &=  \left(1 - \frac{2y_1}{3}\right)(y_2-y_1)\xi\,, \nonumber\\
    \Omega_{\text{geo}} & =  1 - (y_2 - y_1)\xi\,, \label{Omegas_second_case}
\end{align}
and
\begin{align}
    w_{\text{eff}} = -\frac{2y_1}{9} + \frac{1}{3} -\frac{1}{1-\frac{f''R}{f'}} \left[ \frac{(3-y_1)f''R}{3f'} + \frac{\xi'}{\xi}\right]. \label{weff_sec_case}
\end{align}
Eqs.~(\ref{Omegas_second_case}) and (\ref{weff_sec_case}) are valid for both values in Eq.~(\ref{y_1_second_case}).

As discussed previously, $\Omega_{\text{geo}}$ can behave differently on different epochs. The critical points satisfying Eq.~(\ref{Omegas_y1_3_first_case}) will have $\Omega_{\text{geo}}$ standing for a dark energy ($w_{\text{geo}} = w_{\text{eff}} = -1$, by Eq.~(\ref{individual_w})), which will be the unique constituent of the universe and, by Eq.~(\ref{acceleration_eq}), will describe an accelerated period of expansion of a de Sitter kind, since the scale factor will be an exponential function of time. But on the critical points satisfying Eq.~(\ref{Omegas_y1_0_first_case}), $\Omega_{\text{geo}}$ behaves effectively as a radiation fluid, because, by Eq.~(\ref{individual_w}) again, $w_{\text{geo}}=1/3$. In this case, the universe is in a decelerated expansion stage.   
One must note, however, that, in principle, some $f(R)$ theories may not have critical points of these kinds, since this conclusion depends, at least, on Eq.~(\ref{fixed_points_condition_1}) having a real solution. 

We can obtain further information from the critical points of Eq.~(\ref{Omegas_y1_3_first_case}) by calculating the matrix of partial derivatives of the RHS of Eqs.~(\ref{y_1_evolution}) and (\ref{y_2_evolution}) with respect to $y_1$ and $y_2$. The matrix expression for a general $f(R)$ is given in Appendix \ref{app:Jacobian_matrix}, together with its value in the critical points. From these results, and assuming no divergent term is present, it is straightforward to conclude that the eigenvalues of the critical points of the form $(3,y_2)$ are negative (the same values encountered for $\mathcal{P}_5$ in Table \ref{tab:fixed_points_alpha_dif_1}) for a wide variety of Palatini $f(R)$ theories. They are, then, attractors. 

The following subsections are focused on exponential gravity (cf. Eq.~(\ref{exp_gravity})). The study of the phase space of this particular theory has two main objectives: explore the qualitative behavior of its solutions, since for the Palatini formalism this has not yet been done in the literature, and exemplify the application of the new approach presented in this paper to an $f(R)$ theory for which the FTT method cannot be fully employed, highlighting possible subtleties that may emerge throughout the analysis. We will discuss the case $\alpha = 1$ separately because it needs a little more caution. 

\subsection{Exponential gravity with $\alpha \neq 1$}

In the case of exponential gravity, given by Eq.~(\ref{exp_gravity}), the function $f/R$ is a strictly increasing function of $R$. Because of this, the second definition in Eq.~(\ref{new_variables_def}) can always be solved for $R$ as a function of $y_2-y_1$. This implies that our approach can be used to obtain a closed and autonomous dynamical system.

We begin by describing the domain and physically forbidden zones in the phase space $(y_1,y_2)$. As discussed previously, the locus of points where $R$ has a fixed value are the straight lines with inclination of 45 degrees and passing through the point $(0, f/R)$. By making the limits $R \rightarrow \pm \infty$ in Eq.~(\ref{exp_gravity}), we find:
\begin{align}
    -\infty < \frac{f}{R} < 1.
\end{align}
The straight line $R= \infty$ contains, then, the point $(0,1)$. Since $f/R$ is an increasing function of $R$, all the points above this line would belong to a line with a greater value of $R$, but there cannot be a greater value than $\infty$, from which we conclude that these points are out of the domain. The line $R=0$ has the point $(0,1-\alpha)$. Below it, $R<0$ and above it, $R>0$.

Since $\xi > 0$ and $H^2>0$, the first definition of Eq.~(\ref{new_variables_def}) implies that the sign of $y_1$ must be the same as the ratio $f/f'$. This imposes further restrictions on $y_1$. It can be shown that, for $\alpha>1$, the solution $R_0$ of the equation $f(R)=0$ is positive and that $f/f'$ is non-positive in the intervals of $x(R)\leqdef R/R_{\star}$ given by $(-\infty,0)$ and $(\ln \alpha,x(R_0))$. For $\alpha<1$, $f(R)$ vanishes in a negative value $\bar{R}_0$ and $f/f'$ is non-positive on the intervals $(-\infty, x(\bar{R}_0))$ and $(\ln \alpha,0)$. After restricting $y_1$, the second definition of Eq.~(\ref{new_variables_def}) imposes no additional constraint on $y_2$.

\begin{figure*}
	\centering
	\includegraphics[scale=0.36]{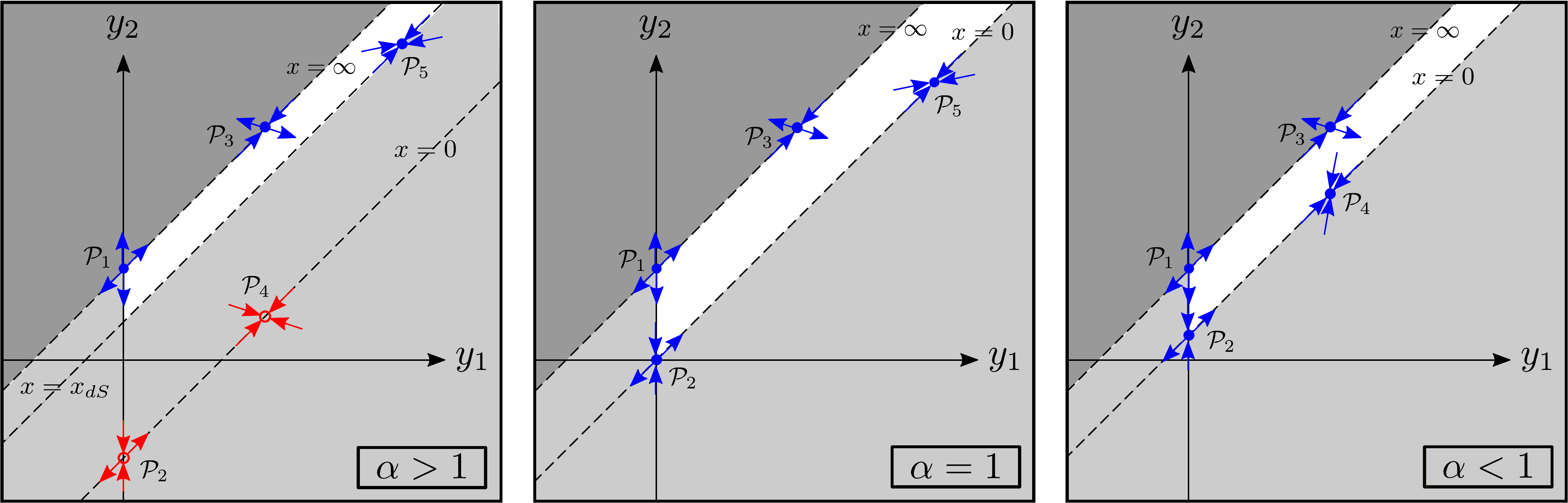}
	\caption{Phase space diagram for exponential gravity illustrating the physically acceptable (filled blue) and unacceptable (empty red) critical points and the associated eigenvectors. Light grey regions do not correspond to physically admissible solutions of the original background equations since they have either $\rho_m < 0$ and/or $H^2 < 0$ and/or $\Omega_r < 0$. The dark grey region is outside of the domain of the variables $y_1$ and $y_2$.}
	\label{fig:phase_space}
\end{figure*}

Finally, we assume, for physical reasons, that $\rho_m$ is never negative, turning Eq.~(\ref{trace}) into a restriction on the values of $R$ which, in $\Lambda$ Cold Dark Matter ($\Lambda$CDM), gives $R \geq 4\Lambda$. In contrast, for exponential gravity, if $\alpha < 1$, the RHS of Eq.~(\ref{trace}) is non-negative for any $R$. On the other hand, for $\alpha>1$, this constraint eliminates the region between the lines $x=0$ and $x=x_{dS}$, where $x_{dS}$, as we will see next, is the positive root of Eq.~(\ref{fixed_points_condition_1}), while the lines themselves remain to be physically acceptable ($\rho_m = 0$ on them). The physically acceptable domain is presented in Fig.~(\ref{fig:phase_space}) as the white regions.

The next step is to obtain the roots of Eq.~(\ref{fixed_points_condition_1}). They are two, $x_{dS} > 0$ and $x_{*} < 0$, but exist only for $\alpha > 1$ and can be found numerically (point $\mathcal{P}_5$ on the first panel of Fig.(\ref{fig:phase_space}) correspond to $x = x_{dS}$). We have already concluded that, in principle, for $x_{dS}$ and $x_{*}$, the possible values of $y_1$ are given by Eq.~(\ref{y_1_first_case}). If $y_1 = 0$, the first of Eq.~(\ref{new_variables_def}) implies that either $f = 0$ or $H^2f'\xi \rightarrow \pm \infty$. However, the former is not true for both solutions and the latter would lead to $\rho_r \rightarrow \infty$, from Eq.~(\ref{Omega_r}), which implies $a \rightarrow 0$ and, then, $\rho_m \rightarrow \infty$ in view of Eqs.~(\ref{matter_conservation}) and (\ref{radiation_conservation}). But from Eq.~(\ref{trace}) we find $\rho_m = 0$ at $x_{dS}$ and $x_{*}$. Therefore the critical points that are solutions to Eq.~(\ref{fixed_points_condition_1}) can only have $y_1 = 3$, for exponential gravity. This implies that the critical point with $x_{*}<0$ has a positive $y_1$. This is, for $\alpha>1$, as previously discussed, a forbidden region of the phase space and, thus, does not correspond to a real solution of the original differential system, so we will neglect it. The corresponding value $y_{2,dS}$ for the root $x_{dS}$ can be found numerically by the second definition of Eq.~(\ref{new_variables_def}).  Lastly, as pointed out before, we note that the value of $x_{dS}$ together with Eq.~(\ref{Omegas_y1_3_first_case}) in fact allows us to interpret this fixed solution as a de Sitter universe.

The remaining critical points are the solutions of Eq.~(\ref{fixed_points_condition_2}), which implies either $R = 0$ or $R \rightarrow \infty$. For both values, the first of Eq.~(\ref{y_1_second_case}) gives $y_1 = 3/2$. Putting this into Eq.~(\ref{fixed_points_condition_2}) again we find
\begin{align}
    y_2 = 
    \begin{cases}
    \frac{5}{2}\,, \quad \text{for} \; R \rightarrow \infty \\
    \frac{5}{2} - \alpha\,, \quad \text{for} \; R = 0\,.
    \end{cases}
\end{align}
Lastly, when $y_1 = 0$, 
\begin{align}
    y_2 = 
    \begin{cases}
    1\,,  \quad \text{for} \; R \rightarrow \infty\\
    1 - \alpha\,, \quad \text{for } R = 0\,.
    \end{cases}
\end{align}
By replacing the coordinates of these four critical points, together with the corresponding values of $R$, into Eqs.~(\ref{Omegas_second_case}) and (\ref{weff_sec_case}) we find the values for $w_{\text{eff}}$ and the $\Omega's$ that are shown in Table \ref{tab:fixed_points_alpha_dif_1}. Overall we obtained one regular and one effective radiation dominated phase (points $\mathcal{P}_1$ and $\mathcal{P}_2$, respectively), one regular and one effective dust dominated phase (points $\mathcal{P}_3$ and $\mathcal{P}_4$, respectively) and, for $\alpha>1$, one physically admissible dark energy dominated epoch (points $\mathcal{P}_5$). As can be verified using Eq.~(\ref{individual_w}), the geometrical component behaves effectively as radiation in $\mathcal{P}_2$, as dust in $\mathcal{P}_4$ and as dark energy in $\mathcal{P}_5$. It is important to emphasize that, for $\alpha<1$ there is no de Sitter critical point. 

Furthermore, $H^2$ can be determined as well. For critical points $\mathcal{P}_1$ and $\mathcal{P}_2$, we note by Eq.~(\ref{trace}) that $\rho_m = \infty$ and $\rho_m = 0$, respectively. Because of Eqs.~(\ref{matter_conservation}) and (\ref{radiation_conservation}), one finds $\rho_r = \rho_m$ in these points. Replacing these values in Eq.~(\ref{Hubble_Param}) and making the appropriate limit in $R$ for both critical points, one finds $H^2 = \infty$ for $\mathcal{P}_1$ and $H^2= 0$ for $\mathcal{P}_2$. For points $\mathcal{P}_3$ and $\mathcal{P}_4$, one can use the definition for $y_1$ in Eq.~(\ref{new_variables_def}), obtaining $H^2 = \infty$ and $H^2= 0$ respectively. Finally, for point $\mathcal{P}_5$, a similar procedure is done, but Eq.~(\ref{fixed_points_condition_1}) is used as well. This results on the numerical values for $H^2$ found on Table \ref{tab:fixed_points_alpha_dif_1}.

Additionally, it is possible to obtain, by the results presented in Appendix \ref{app:Jacobian_matrix}, the eigenvalues of the Jacobian matrix in each critical point, which are shown in Table \ref{tab:fixed_points_alpha_dif_1} together with the critical point character (attractor, repeller or saddle).

\subsection{Exponential gravity with $\alpha = 1$}

The domain in this case is still restrained, by the same reasons as before, to the area below the line where $R$ diverges. The sign of $y_1$ continues to be given by the sign of $f/f'$, but now we simply find that it must be positive for $R>0$ and negative for $R<0$. Furthermore, the region where $\rho_m<0$ corresponds to $R<0$, and it must be discarded for physical reasons. 

The $\alpha = 1$ case has to be analyzed separately because of apparent divergences. First, Eq.~(\ref{fixed_points_condition_1}) is only satisfied when $R = 0$, but then the denominator appearing in the dynamical equations, $f' - f''R$, also vanishes and so some results previously obtained for a generic $f(R)$ must be reassessed. Eq.~(\ref{critical_condition}) is still satisfied in the limit $R \rightarrow 0$, which implies, by the second of Eqs.~(\ref{new_variables_def}), that if a pair $(y_1,y_2)$ makes the RHS of any of the Eqs.~(\ref{y_1_evolution}) or (\ref{y_2_evolution}) vanish, it is a critical point. Replacing $y_1-y_2$ by $f/R$ in Eq.~(\ref{y_2_evolution}), taking the limit of $R\rightarrow0$ on it, noting that 
\begin{align}
    &\lim_{R\to0}\frac{f' - 2f/R}{f'-f''R} = -\frac{1}{3},
\end{align}
while, in the same limit, $f/R \to 0$, $f'R/f \to 2$ and $f''R^2/f \to 2$ and imposing $dy_2/dN = 0$ we find
\begin{align}
    y_1 = y_2 = 0 \quad \text{or} \quad 3\,.
\end{align}

Furthermore, Eq.~(\ref{fixed_points_condition_2}) has two roots: $R = 0$ and $R \rightarrow \infty$. By making a limiting procedure analogous to the previous one, but now on Eq.~{\ref{y_1_evolution}} and imposing $dy_1/dN = 0$ we get
\begin{align}
y_1 =
\begin{cases}
    0 \quad \text{or}\quad 3\,, \quad \text{for} \; R=0\,, \\
    0 \quad \text{or} \quad \frac{3}{2}\,, \quad \text{for} \; R \rightarrow \infty\,.
\end{cases}
\end{align}
Table \ref{tab:fixed_points_alpha_eq_1} summarizes the critical points for the $\alpha = 1$ model and also shows the values of the $\Omega$'s and $w_{\text{eff}}$
at each of them. As in the $\alpha >1$ case, here there is a dark energy point (point $P_5$), but it has $w_{\text{eff}} = -2/3$. Almost all values of $H^2$ can be obtained by the same methods used on the $\alpha\neq 1$ case, with the exception of point $\mathcal{P}_2$. In this point, one has to make the limit $R\to0$ in Eq.(\ref{Hubble_evolution}) and impose that the derivative of $H^2$ vanishes (since this must be a critical point of the original dynamical system), finding that $H$ must vanish as well. The $\alpha = 1$ model clearly defines a discontinuity on the parameter space of exponential gravity, since, for example, the dust dominated attractor (point $\mathcal{P}_4$) disappears. Again, in Appendix \ref{app:Jacobian_matrix} we make explicit the Jacobian matrices at the fixed points and their corresponding eigenvalues are shown in Table \ref{tab:fixed_points_alpha_eq_1}.

\begin{table*}[t] 
	\centering
	\caption{Critical points for $\alpha \neq 1$}
	\label{tab:fixed_points_alpha_dif_1}
	\setlength{\extrarowheight}{5pt}
	\begin{tabular}{|c|c|c|c|c|c|c|c|c|c|}
		\hline
		model & $(y_1, y_2)$ & $w_{\text{eff}}$ & $\Omega_m$ & $\Omega_r$ & $\Omega_{\text{geo}}$ & $x$ & $H^2$ & eigenvalues & type \\ \hline
		\multirow{2}{*}{$\alpha > 1$} & $\mathcal{P}_1 : (0, 1)$ & 1/3 & 0 & 1 & 0 & $\infty$ & $\infty$ & 1, 3 & repeller \\ \cline{2-10}
		\multirow{2}{*}{or} & $\mathcal{P}_2 : (0, 1-\alpha)$ & 1/3 & 0 & $1-\alpha$ & $\alpha$ & 0 & 0 & 1, $-3$ & saddle \\ \cline{2-10}
		\multirow{2}{*}{$\alpha < 1$} & $\mathcal{P}_3 : (3/2, 5/2)$ & 0 & 1 & 0 & 0 & $\infty$ & $\infty$ & $-1$, 3 & saddle \\ \cline{2-10}
		& $\mathcal{P}_4 : (3/2, 5/2-\alpha)$ & 0 & $1-\alpha$ & 0 & $\alpha$ & 0 & 0 & $-1$, $-3$ & attractor \\ \hline
		\multirow{1}{*}{only $\alpha>1$} & $\mathcal{P}_5 : \left(3, y_{2,dS}(\alpha)\right)$ & $-1$ & 0 & 0 & 1 & $x_{dS}(\alpha) > 0$ & $R_{\ast}x_{dS}/12$ & $-3$, $-4$ & attractor\\ \hline
	\end{tabular}
\end{table*}

\begin{table*}[t] 
	\centering
	\caption{critical points for $\alpha = 1$}
	\label{tab:fixed_points_alpha_eq_1}
	\setlength{\extrarowheight}{5pt}
	\begin{tabular}{|c|c|c|c|c|c|c|c|c|}
		\hline
		$(y_1, y_2)$ & $w_{\text{eff}}$ & $\Omega_m$ & $\Omega_r$ & $\Omega_{\text{geo}}$ & $x$ & $H^2$ & eigenvalues & type \\ \hline
		$\mathcal{P}_1 : (0, 1)$ & 1/3 & 0 & 1 & 0 & $\infty$ & $\infty$ & 1, 3 & repeller \\ \hline
		$\mathcal{P}_2 : (0, 0)$ & 0 & 0 & 0 & 1 & 0 & 0 & $-1$, 2& saddle\\ \hline
		$\mathcal{P}_3 : (3/2, 5/2)$ & 0 & 1 & 0 & 0 & $\infty$ & $\infty$ & $-1$, 3 & saddle \\ \hline
		$\mathcal{P}_5 : (3, 3)$ & -2/3 & 0 & 0 & 1 & 0 & 0 & $-1$, $-2$ & attractor \\ \hline
	\end{tabular}
\end{table*}

\begin{figure*}
\centering
\includegraphics[width=1.0\textwidth]{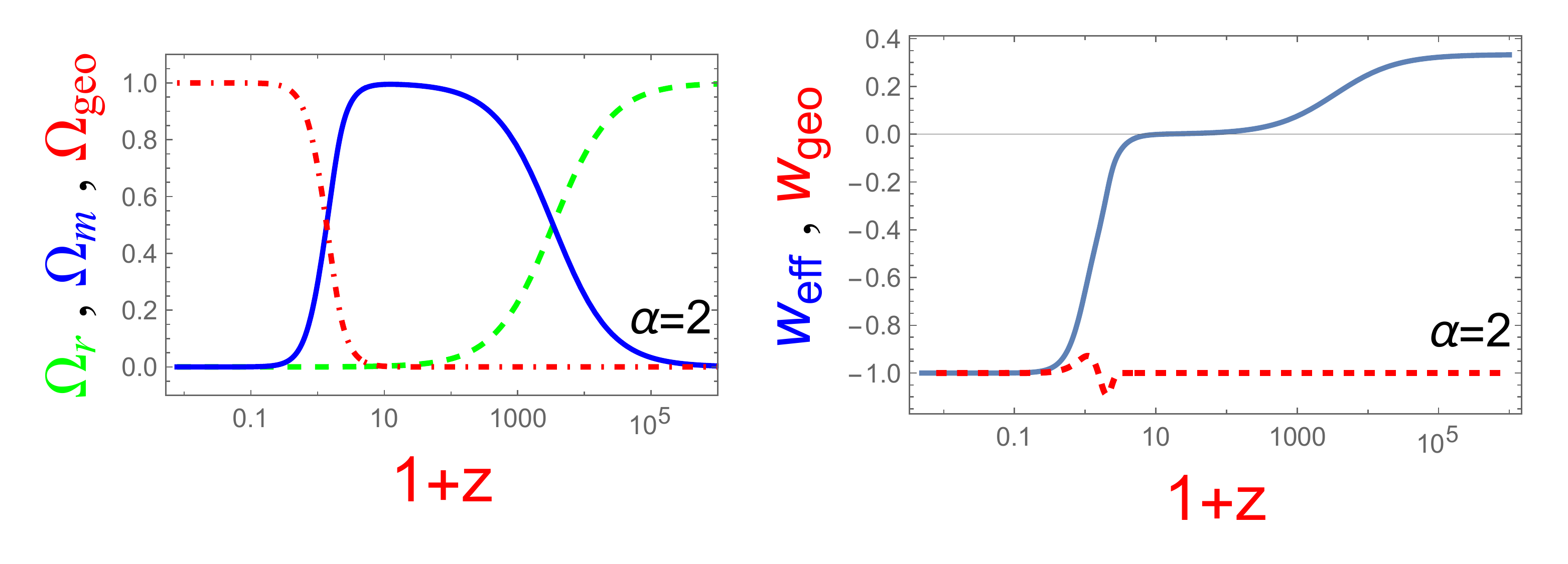}%
\caption{\small{ In the left panel we show the evolution of the energy density parameters  [$\Omega_{r}$ (green dashed), $\Omega_m$ (blue), $\Omega_{geo}$(red dot-dashed)] and in the right panel we show the effective equation of state parameter (blue) and the equation of state parameter of the geometric component (red dashed), for a typical solution in exponential gravity with $\alpha = 2$ which starts in the radiation dominated era, with $\tilde{\Omega}_{m0} = 0.3$. In this case there is a regular matter dominated phase and the system follows the points $\mathcal{P}_1$, $\mathcal{P}_3$ and the final de Sitter attractor $\mathcal{P}_5$ with $w_{\text{eff}}=-1$.}}
\label{fig:omega_weff_alpha2}
\end{figure*}

\begin{figure*}
\centering
\includegraphics[width=1.0\textwidth]{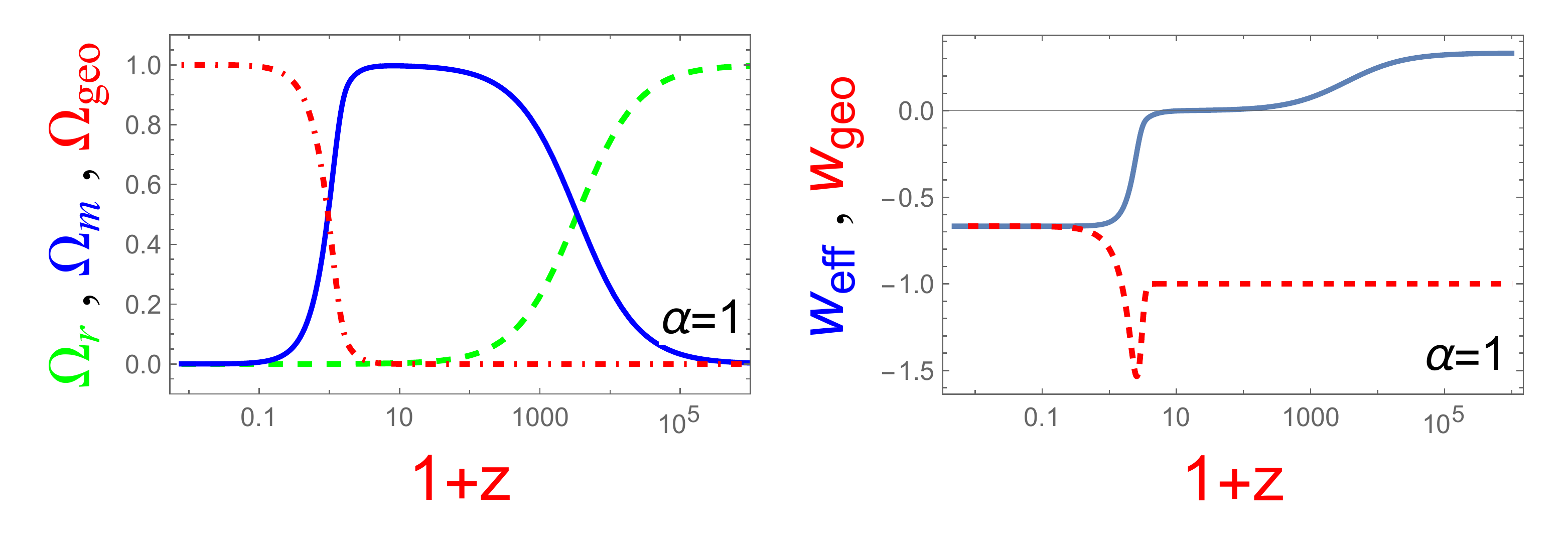}%
\caption{\small{ In the left panel we show the evolution of the energy density parameters  [$\Omega_{r}$ (green dashed), $\Omega_m$ (blue), $\Omega_{geo}$(red dot-dashed)] and in the right panel we show the effective equation of state parameter (blue) and the equation of state parameter of the geometric component (red dashed), for a typical solution in exponential gravity with $\alpha = 1$ which starts in the radiation dominated era, with $\tilde{\Omega}_{m0} = 0.3$. In this case there is a regular matter dominated phase and the system follows the points $\mathcal{P}_1$, $\mathcal{P}_3$ and the final attractor $\mathcal{P}_5$ with $w_{\text{eff}}=-2/3$.}}
\label{fig:omega_weff_alpha1}
\end{figure*}

\begin{figure*}
\centering
\includegraphics[width=1.0\textwidth]{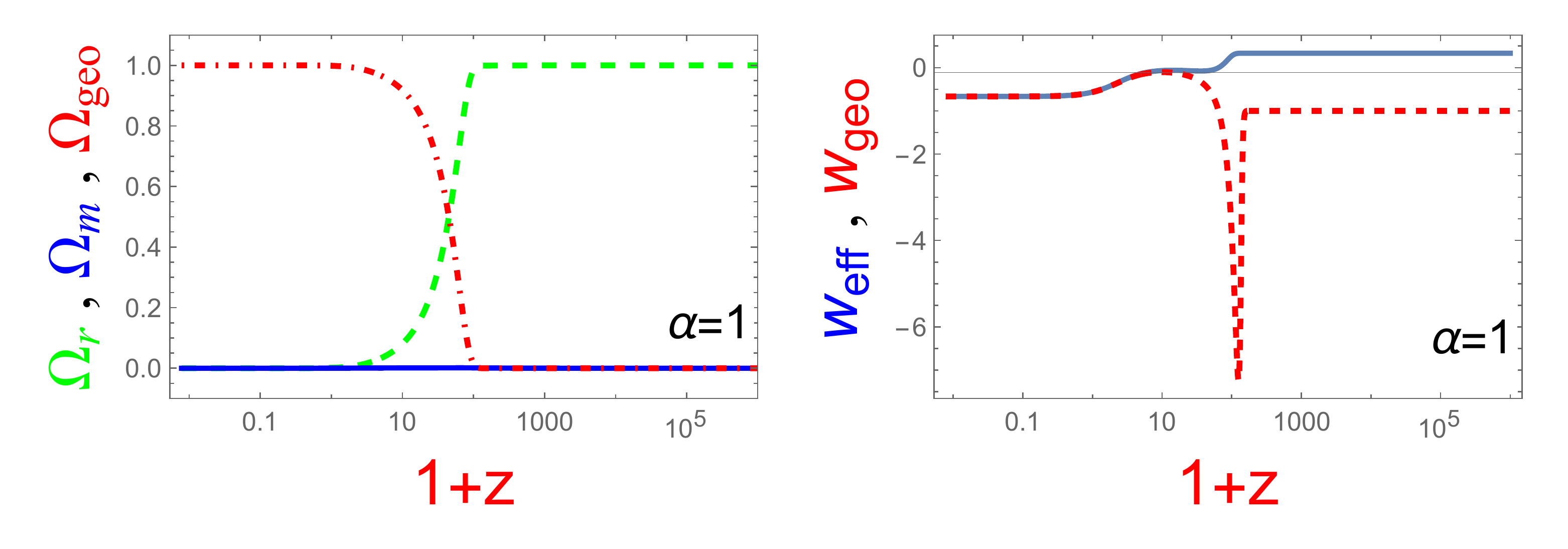}%
\caption{\small{In the left panel we show the evolution of the energy density parameters  [$\Omega_{r}$ (green dashed), $\Omega_m$ (blue), $\Omega_{geo}$(red dot-dashed)] and in the right panel we show the effective equation of state parameter (blue) and the equation of state parameter of the geometric component (red dashed), for a typical solution in exponential gravity with $\alpha = 1$ which starts in the radiation dominated era, with $\tilde{\Omega}_{m0} = 10^{-5}$. In this case there is not a regular matter dominated phase and the system follows the points $\mathcal{P}_1$, $\mathcal{P}_2$ (with $w_{\text{eff}}=0$) and the final attractor $\mathcal{P}_5$ with $w_{\text{eff}}=-2/3$.}}
\label{fig:P1P2P5}
\end{figure*}

\begin{figure*}
\centering
\includegraphics[width=1.0\textwidth]{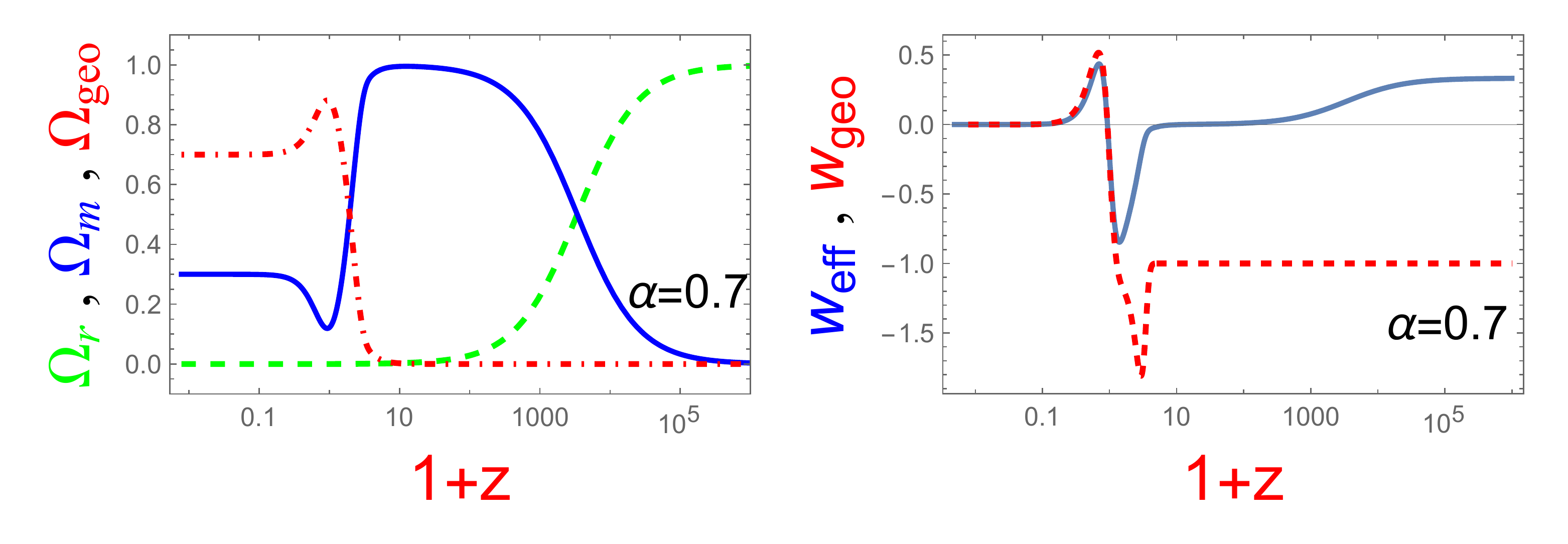}
\caption{\small{ In the left panel we show the evolution of the energy density parameters  [$\Omega_{r}$ (green dashed), $\Omega_m$ (blue), $\Omega_{geo}$(red dot-dashed)] and in the right panel we show the effective equation of state parameter (blue) and the equation of state parameter of the geometric component (red dashed), for a typical solution in exponential gravity with $\alpha = 0.7$ which starts in the radiation dominated era, with $\tilde{\Omega}_{m0} = 0.3$. In this case there is a regular matter dominated phase and the system follows the points $\mathcal{P}_1$, $\mathcal{P}_3$ and the final attractor $\mathcal{P}_4$ with $w_{\text{eff}}=0$.}}
\label{fig:omega_weff_alpha07}
\end{figure*}

\begin{figure*}
\centering
\includegraphics[width=1.0\textwidth]{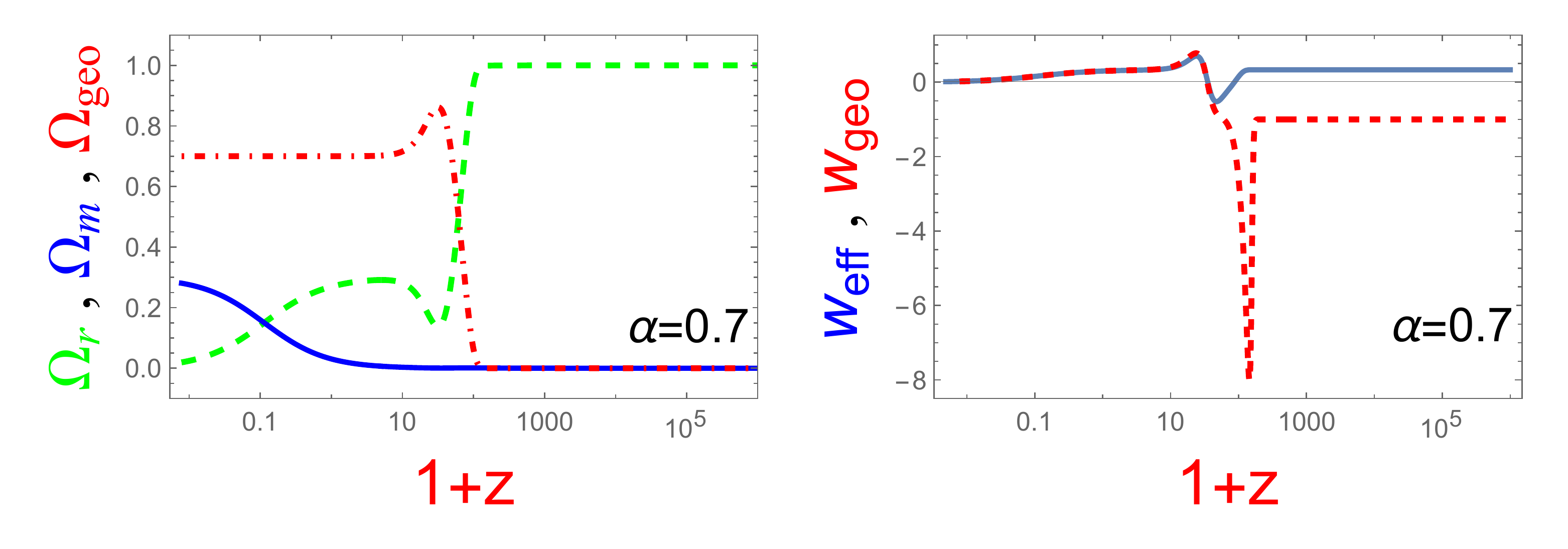}
\caption{\small{ In the left panel we show the evolution of the energy density parameters  [$\Omega_{r}$ (green dashed), $\Omega_m$ (blue), $\Omega_{geo}$(red dot-dashed)] and in the right panel we show the effective equation of state parameter (blue) and the equation of state parameter of the geometric component (red dashed), for a typical solution in exponential gravity with $\alpha = 0.7$ which starts in the radiation dominated era, with $\tilde{\Omega}_{m0} = 10^{-5}$. In this case, there is not a regular matter dominated phase and the system follows the points $\mathcal{P}_1$, $\mathcal{P}_2$ (with $w_{\text{eff}}=1/3$) and the final attractor $\mathcal{P}_4$ with $w_{\text{eff}}=0$.}}
\label{fig:P1P2P4}
\end{figure*}

\section{Discussion} \label{sec:discussion}

Following Fig.~\ref{fig:phase_space}, we are able to understand the viable behavior of solutions qualitatively. The arrows in the figure indicate the direction of the eigenvectors in each critical point. 

For $\alpha>1$, $\mathcal{P}_4$ is an isolated point, it can only be  achieved by a solution which has $R=0$ throughout its evolution, which is not physically admissible, since by Eq.(\ref{trace}), it would have $\rho_m = 0$ and, consequently by Eq.(\ref{matter_conservation}) and the fact that $\rho_{m0}\neq 0$ in our analysis, the scale factor would be divergent. Point $\mathcal{P}_2$ is not physically admissible as well for $\alpha>1$. This is because, by Eq.(\ref{Omega_r}), $\Omega_r$ is negative when approaching this point by negative values of $R$ with $y_1<0$ and, by approaching in any other way, either $\rho_m<0$ or $H^2<0$. The other three points, $\mathcal{P}_1$, $\mathcal{P}_3$ and $\mathcal{P}_5$ are respectively radiation, matter and dark energy dominated and provide the expected qualitative evolution of our observed universe. Giving an initial condition near the radiation dominated critical point, the solution will be repelled until it reaches the matter dominated era, which is a saddle point. Then, it will continue to evolve until it reaches the final attractor of a dark energy dominated era, tending to it asymptotically. Almost all solutions tend to the attractor through the direction of the eigenvector that is not parallel to the $x = x_{dS}$ line, since this direction is related with the eigenvalue of lowest absolute value, so that the Ricci scalar reaches $R_{dS}$ only asymptotically. The $\alpha$-dependent eigenvectors are only those of point $\mathcal{P}_4$ that are not parallel to the $R=0$ line and of point $\mathcal{P}_5$ which is not aligned with the $x=x_{dS}$ line.

It is important to emphasize that our approach is complementary to the traditional FTT one, since both assume invertibility of certain functions of the Ricci scalar, which may fail depending on the theory of interest. In the case where both approaches fail, we suggest that a new approach suitable for the specific $f(R)$ may be found by constructing an invertible function of $R$, call it $g(R)$, and redefining the variable $y_2$ in Eq.~(\ref{new_variables_def}) by exchanging $f/R$ for $g(R)$, without changing the definition of $y_1$.

Still for $\alpha>1$, one could argue that the FTT approach is valid, since the LHS of Eq.~(\ref{FTT_constraint}) is an invertible function of $R$. But if this approach was used to investigate this case, one would find points $\mathcal{P}_1$ and $\mathcal{P}_2$ corresponding to the same pair $(\bar{y}_1, \bar{y}_2)$. This would occur for $\mathcal{P}_3$ and $\mathcal{P}_4$ as well. This coincidence in the FTT approach would indicate ambiguity when analysing the behavior of solutions near these critical points, since a repeller would coincide with a saddle point and an attractor with another saddle point. Additionally, the non-physical nature of $\mathcal{P}_2$ and $\mathcal{P}_4$ because of the $\rho_m<0$ region that separates them from the physical points would not be evident, as it is with our new approach. These points coincide in the FTT variables because $\bar{y}_1 = 0$ on them and so the RHS of Eq.~(\ref{FTT_constraint}) is not well defined, making it impossible to obtain $R$ as a function of such variables in these points by this relation.

The usefulness of introducing new variables becomes clear by the same reasons. If we were to analyze the dynamical system in the $(R,H^2)$ plane, multiple critical points would be represented as the same point on this phase space, as Tables \ref{tab:fixed_points_alpha_dif_1} and \ref{tab:fixed_points_alpha_eq_1} show. Also, there would be important critical points existing on infinity, which would require a more careful asymptotic analysis. By using the new variables $y_1$ and $y_2$, we solve both of these problems.

In \cite{Bamba2010}, a numerical analysis was made for the exponential gravity in the metric formalism, for values of $\alpha>1$. The cosmological evolution of the density parameters $\Omega_m$, $\Omega_r$ and $\Omega_{\text{geo}}$ are very similar to what we have found in our work for this case, as well as the behavior of $w_{\text{geo}}$ as a function of redshift. The redshift where the equality between dust and the geometrical component takes place is, in both cases, of order unity.

For $\alpha = 1$, there are no isolated points. Points $\mathcal{P}_1$, $\mathcal{P}_3$ and $\mathcal{P}_5$ play the same role as before, but there are solutions that can pass near point $\mathcal{P}_2$, an intermediate stage of effective dust domination with low $R$, where the geometrical component dominates and behaves as dust. One can show that the RHS of Eq.~(\ref{ricci_evolution}) is always negative for exponential gravity, which implies that $R$ must always decrease with $N$ in a given solution. As a consequence, no solution can go from $\mathcal{P}_2$ to $\mathcal{P}_3$. The two possible behaviors for the solutions when giving an initial condition near $\mathcal{P}_1$ are: either it passes by $\mathcal{P}_2$ or by $\mathcal{P}_3$, always ending in $\mathcal{P}_5$. Both intermediate stages are regularly or effectively dust dominated and with a saddle-like nature, differing only by the value of $R$ and $H^2$ and by how relevant the geometrical component is. The dark energy epoch on $\mathcal{P}_5$ is different from the one in the case $\alpha>1$, since the fluid satisfies $w_{geo} = -2/3$ instead of $w_{geo} = -1$, and thus the accelerated expansion is not of a de-Sitter kind.

Finally, for $\alpha<1$, some non-physical points of the phase space for $\alpha>1$ become physical. The big difference in this case is that the final attractor, point $\mathcal{P}_4$, is dominated by an effective dust coming from the mixture of regular dust and geometrical component and there is no dark energy dominated era, with a steady accelerated expansion, only a possible transient period of this kind, as Fig.~(\ref{fig:omega_weff_alpha07}) suggests. Because of the always decreasing value of the Ricci scalar along a given solution, one can have, in principle, for a solution starting near $\mathcal{P}_1$, an intermediate state of radiation dominated kind and low $R$ (point $\mathcal{P}_2$), where the fluid is a mixture of regular radiation and geometrical component, or it can have a regular dust dominated intermediate stage (point $\mathcal{P}_3$), both always ending on the attractor $\mathcal{P}_4$. Almost all solutions in this case tend to $\mathcal{P}_4$ through the direction of the $R=0$ line, and the direction of the other pair of eigenvectors is $\alpha$-dependent. 

We now show typical numerical solutions 
of the background equations in exponential gravity. Figs. \ref{fig:omega_weff_alpha2}, \ref{fig:omega_weff_alpha1}, \ref{fig:P1P2P5}, \ref{fig:omega_weff_alpha07} and \ref{fig:P1P2P4} show the evolution of the density parameters, the effective equation of state parameter and the geometric equation of state parameter for solutions starting in the radiation dominated era. For $\alpha > 1$, Fig. \ref{fig:omega_weff_alpha2}, we see that the solution begins near the repeller $\mathcal{P}_1$, evolves towards the saddle matter dominated point $\mathcal{P}_3$ and ends at the de Sitter attractor $\mathcal{P}_5$. For $\alpha < 1$ and $\tilde{\Omega}_{m0}=0.3$, Fig. \ref{fig:omega_weff_alpha07}, the main difference is that the final attractor era, $\mathcal{P}_4$, behaves effectively like a dust dominated fluid
with both matter and the geometrical component. For $\alpha < 1$ and $\tilde{\Omega}_{m0}=10^{-5}$, Fig. \ref{fig:P1P2P4}, there is no regular matter dominated phase and the system follows the points $\mathcal{P}_1$, $\mathcal{P}_2$ and again the final attractor $\mathcal{P}_4$ with $w_{\text{eff}}=0$. For $\alpha =1$ and $\tilde{\Omega}_{m0}= 10^{-5}$, Fig. \ref{fig:P1P2P5} the system begins near the repeller $\mathcal{P}_1$, evolves towards the saddle point $\mathcal{P}_2$ (with $w_{\text{eff}}=0$) and ends at the final attractor $\mathcal{P}_5$ with $w_{\text{eff}}=-2/3$. Finally, for $\alpha = 1$ and $\tilde{\Omega}_{m0}= 0.3$, Fig. \ref{fig:omega_weff_alpha1}, there is a matter dominated era and the final phase is again dominated by the geometrical component alone with $w_{\text{eff}} = -2/3$; this particular solution never crosses the neighborhood of the saddle point $\mathcal{P}_2$. The numerical solutions for all cases are, then, compatible with the results from the qualitative analysis. We remark that the above $\tilde{\Omega}_{m0}$ denotes the present value of the matter density parameter that a $\Lambda$CDM model would have, if it had the same present matter density as the exponential gravity $f(R)$ model \cite{Matos2021}.

As discussed above in Palatini exponential gravity it is possible to have the following consecutive three phases: a radiation dominated era, a matter dominated era and a late time acceleration phase.
It is clear that the occurrence of these phases {\it per se} does not guarantee the viability of the model. It is also necessary that, at least for some values of the model parameters, it is in accordance with observations.
In exponential gravity, as we increase $\alpha$, the background behavior of the model approaches that of $\Lambda$CDM with the equation of state parameter of the
geometric component ($w_{geo}$) becoming more and more close to the value $w_{geo}=-1$ for all values of the redshift. Therefore, since  $\Lambda$CDM is in good agreement with observations, we expect that background tests, those that essentially depend only on distances, like SNeIa and BAO for instance, should impose relatively large values for the parameter $\alpha$. In fact, more restrictive
then the background tests are those that depend on the matter density perturbations. It can be shown that, in Palatini $f(R)$, the linear matter perturbations satisfy the following approximate equation \cite {Tsujikawa2007,DeFelice2010}:

\begin{equation}
\ddot{\delta}_m+2H\dot{\delta}_m-\frac{\kappa \rho_m}{2f'}\left[1+\frac{mk^2/(a^2R)}{1-m} \right]\delta_m \simeq 0,
\label{delf}
\end{equation}
where $m \leqdef R f''/f'$. In GR, $m=0$, and there is no scale dependence for the density contrast in the linear regime. For $\Lambda$CDM, the growing mode can be expressed in terms of hypergeometric functions $_2F_1$ as  \cite{Silveira1994}
\begin{equation}
\tilde{\delta}_{m} \propto \frac{1}{1+z} \;{_2}F_1\left[  \frac{1}{3},1, \frac{11}{6};- \frac{1-\tilde{\Omega}_{m0}}{(1+z)^{3}\tilde{\Omega}_{m0}}\right]. 
\label{delw}
\end{equation}
We solved Eq. (\ref{delf}) numerically and obtained the growing mode for exponential gravity. By using Eq.~(\ref{delw}), we then obtained the fractional change in the linear matter power spectrum $P(k)$ relative to $\Lambda$CDM, $\Delta P_k/P_k$, at $z=0$. By assuming $\tilde{\Omega}_{m0}=0.3$ and imposing that the fractional change in the matter power spectrum at $z=0$ cannot be higher than $0.2$ at $k = 0.1 \; (h Mpc^{-1})$, we obtain the constraint $\alpha>5.26$. A complete analysis taking into account data at all scales of the linear mass power spectrum should be even more restrictive, possibly requiring larger values of the parameter $\alpha$. For such high values of $\alpha$, the background evolution of exponential gravity cannot be discriminated from $\Lambda$CDM.

\paragraph*{Acknowledgments.} ---
 We thank Sérgio Quinet de Oliveira for useful discussions. J. C. L. thanks Brazilian funding agency CAPES for
 PhD scholarship 88887.492685/2020-00. I. S. M. thanks Brazilian funding agency CNPq for PhD scholarship GD 140324/2018-6. 
 
 \appendix

\section{Jacobian Matrix}  \label{app:Jacobian_matrix}

The Jacobian matrix is defined by:
\begin{align}
    J_{ab} \leqdef \frac{\partial}{\partial y_b}\left(\frac{dy_a}{dN}\right)\,, \quad a = 1, 2.
\end{align}
By differentiating the RHS of Eqs.~(\ref{y_1_evolution}) and (\ref{y_2_evolution}), one finds, for an arbitrary $f(R)$:
\begin{widetext}
\begin{align}
    J_{11} = & \frac{f' - 2f/R}{f' - f''R}(3f' - y_1f''R)\frac{R^2}{f^2}y_2 - y_1\frac{f''R^2}{f}\frac{f' - 2f/R}{f' - f''R}  + 4 - \frac{2f'R^2}{3f^2}(2y_2 - y_1)y_1 \nonumber \\
    & + y_1\frac{R}{f}\frac{\partial R}{\partial y_1}\bigg\{ - \frac{2y_1f''}{3} + \frac{f'''R(f' - 2f/R)(3f' - y_1f''R)}{(f' - f''R)^2} \nonumber \\& \hspace{55pt}+ \frac{(f'' - 2f'/R + 2f/R^2)(3f' - y_1f''R) +  (f' - 2f/R)[3f'' - y_1(f'''R + f'')]}{f' - f''R}\bigg\} \label{J11}
\end{align}

\begin{align}
    J_{12} = &- \frac{f' - 2f/R}{f' - f''R}(3f' - y_1f''R)\frac{R^2}{f^2}y_1 + \frac{2f'R^2}{3f^2}y_1^2 \nonumber \\
    & + y_1\frac{R}{f}\frac{\partial R}{\partial y_2}\bigg\{ - \frac{2y_1f''}{3} + \frac{f'''R(f' - 2f/R)(3f' - y_1f''R)}{(f' - f''R)^2} \nonumber \\& \hspace{55pt}+ \frac{(f'' - 2f'/R + 2f/R^2)(3f' - y_1f''R) +  (f' - 2f/R)[3f'' - y_1(f'''R + f'')]}{f' - f''R}\bigg\} \label{J12}
\end{align}

\begin{align}
    J_{21} =  &-2 \frac{f' - 2f/R}{f' - f''R}f''\frac{R^2}{f}y_1 - \frac{4f'R}{3f}y_1 + 4 \nonumber \\
    & + \frac{\partial R}{\partial y_1}\bigg\{\frac{1}{f' - f''R}\left[f''-\frac{2f'}{R}+\frac{2f}{R^2} +f'''R\left(f'-\frac{2f}{R}\right) \right]\left[(3f'y_2-y_1^2f''R)\frac{R}{f}-\frac{3f}{R}\right] \nonumber \\& \hspace{35pt} + \frac{f' - 2f/R}{f'-f''R} \left[(3f''y_2 - y_1^2f'''R-y_1^2f'')\frac{R}{f} + (3f'y_2-y_1^2f''R)\frac{f-f'R}{f^2}  - 3\frac{f'R-f}{R^2} \right] \nonumber \\ &\hspace{35pt} -\frac{2y_1^2}{3f} \left[f''R+f'-\frac{f'^2R}{f}\right]\bigg\} \label{J21}
    \end{align}
    
\begin{align}
    J_{22} = & 3\frac{f'R}{f}\frac{f' - 2f/R}{f' - f''R} \nonumber \\
    & + \frac{\partial R}{\partial y_2}\bigg\{\frac{1}{f' - f''R}\left[f''-\frac{2f'}{R}+\frac{2f}{R^2} +f'''R\left(f'-\frac{2f}{R}\right) \right]\left[(3f'y_2-y_1^2f''R)\frac{R}{f}-\frac{3f}{R}\right] \nonumber \\& \hspace{35pt} + \frac{f' - 2f/R}{f'-f''R} \left[(3f''y_2 - y_1^2f'''R-y_1^2f'')\frac{R}{f} + (3f'y_2-y_1^2f''R)\frac{f-f'R}{f^2}  - 3\frac{f'R-f}{R^2} \right] \nonumber \\ &\hspace{35pt} -\frac{2y_1^2}{3f} \left[f''R+f'-\frac{f'^2R}{f}\right]\bigg\} \label{J22}
\end{align}    
\end{widetext}

The derivatives of $R$ can be obtained implicitly by partial differentiating the second of Eqs.~(\ref{new_variables_def}) with respect to $y_1$ and $y_2$, which results in:
\begin{align}
    &\frac{\partial R}{\partial y_1} = \frac{R^2}{f'R-f} = -\frac{\partial R}{\partial y_2}. \label{R_derivatives}
\end{align}

Substituting the above equation on Eqs.~(\ref{J11})--(\ref{J22}), we find the Jacobian matrix on the critical points that satisfy Eq.(\ref{fixed_points_condition_1}) and $y_1=3$ by imposing these conditions additionally and assuming no divergent terms are present:
\begin{align}
    J = A + \left(\frac{6R}{f}-\frac{3f''R^3}{f^2}\right) B, \label{Jacobian_dS}
\end{align}
where 
\begin{align}
    A = \begin{pmatrix}
    -4 && 0 \\ -1 && -3
    \end{pmatrix}
    \end{align}
and
\begin{align}
    B = \begin{pmatrix}
    1 && -1 \\ 1 && -1
    \end{pmatrix}.
\end{align}
By this expression, one is able to conclude that the eigenvalues at these critical points are $-3$ and $-4$, for any $f(R)$ having them.

From now on we develop the necessary results for obtaining the eigenvalues in the case of exponential gravity. The Jacobian of Eq.(\ref{Jacobian_dS}), together with its eigenvalues, is valid for the point $\mathcal{P}_5$. All the remaining critical points of this theory have $R=0$ or $R=\infty$. 

Replacing Eq.~(\ref{R_derivatives}) on Eqs.~(\ref{J11})--(\ref{J22}), taking the limit of $R\rightarrow\infty$ and using that, by the second of Eqs.(\ref{new_variables_def}), $y_2 = y_1 +1$ for this value $R$ we find, for any $\alpha$,
\begin{align}
    J_{exp}(R\rightarrow\infty) = C+\frac{y_1}{3}D,
\end{align}
where
\begin{align}
   C= \begin{pmatrix}
    1 && 0 \\ -2   && 3
    \end{pmatrix}
\end{align}
and 
\begin{align}
   D= \begin{pmatrix}
    5-2y_1 && 2y_1 - 9\\ 5-2y_1   && 2y_1 - 9
    \end{pmatrix}.
\end{align}

 For $\alpha \neq 1$, making the same procedure, but with $R\rightarrow 0$ and $y_2 = y_1 + 1 - \alpha$, the Jacobian matrix becomes:
 \begin{align}
     J_{exp,\alpha\neq1}(R=0) = E + \frac{y_1}{3} F
 \end{align}
 where
 \begin{align}
     E = \begin{pmatrix}
    1 && 0\\ 4 && -3
    \end{pmatrix}
 \end{align}
 and
 \begin{align}
    F = \begin{pmatrix}
     -4 + \frac{9-4y_1}{1-\alpha} && \frac{4y_1-9}{1-\alpha}\\ -4 +  \frac{9-4y_1}{1-\alpha}  && \frac{4y_1-9}{1-\alpha} 
    \end{pmatrix}.
 \end{align}
It is clear that such result is not valid for $\alpha = 1$ because of the matrix $F$. This occurs since, for exponential gravity, $f'-f''R$ vanishes in $R=0$ for $\alpha = 1$, which makes the denominator of several terms of the Jacobian to become 0. The limits of the ratios in which such divergence occurs must be done without separating it in the limits of each part. After performing the limits correctly, one finds:
\begin{align}
     J_{exp,\alpha=1}(R=0) = G + \frac{2y_1}{9}H
 \end{align}
where
\begin{align}
    G = \begin{pmatrix}
    2 && 0\\ 3 && -1
    \end{pmatrix}
\end{align}
and
\begin{align}
     H = \begin{pmatrix}
    y_1-6 && - y_1\\ y_1 - 6 && - y_1
    \end{pmatrix}.
\end{align}

By the values of the Jacobian obtained here, it is trivial to compute the eigenvalues of all the critical points in Tables \ref{tab:fixed_points_alpha_dif_1} and \ref{tab:fixed_points_alpha_eq_1}.

\end{document}